# Smoothed particle hydrodynamics with adaptive spatial resolution (SPH-ASR) for free surface flows

Xiufeng Yang [a, b], Song-Charng Kong [b], Moubin Liu [c], Qingquan Liu [a, *]

[a] School of Aerospace Engineering, Beijing Institute of Technology, Beijing 100081, China

[b] Department of Mechanical Engineering, Iowa State University, Ames, IA 50011, USA

[c] College of Engineering, Peking University, Beijing 100871, China

[*] Corresponding author: liuqq@bit.edu.cn

**Abstract**

A numerical method based on smoothed particle hydrodynamics with adaptive spatial resolution (SPH-ASR) was developed for simulating free surface flows. This method can reduce the computational demands while maintaining the numerical accuracy. In this method, the spatial resolution changes adaptively according to the distance to the free surface by numerical particle splitting and merging. The particles are split for refinement when they are near the free surface, while they are merged for coarsening when they are away from the free surface. A search algorithm was implemented for identifying the particles at the free surface. A particle shifting technique, considering variable smoothing length, was introduced to improve the particle distribution. The presented SPH-ASR method was validated by simulating various free surface flows, and the results were compared to those obtained using SPH with uniform spatial resolution (USR) and experimental data.

# 1. Introduction

Smoothed particle hydrodynamics (SPH), a Lagrangain meshfree particle method, is widely used for simulating various free surface flows in recent years [1, 2], such as dam-break flow [3, 4], water entry [5, 6], liquid sloshing [7, 8], drop impact [9, 10], oil-spill containment [11, 12], and interaction of water and coastal structures [13, 14]. Most of the aforementioned studies using SPH employed uniform spatial resolution (USR). However, when a high spatial resolution is needed, USR will require a large number of SPH particles, which will incur high computational costs such as computation memory and computational time.

If a high spatial resolution is needed only in the region of interest rather than the entire computational domain, it is not necessary to use USR. For example, in mesh-based methods, adaptive mesh refinements are used to resolve the regions of interest, while coarse meshes are used in other regions. Unlike mesh-based methods, in which meshes have fixed connections, the SPH method uses numerical particles to discretize the fluid domain. The SPH particles do not have fixed connections and may move to regions far away from their initial positions. This is originally an advantage of the SPH method, by making it suitable for simulating flows with large deformations or free surfaces. However, the relocation of particles also poses challenges in using non-uniform spatial resolution for SPH simulations, because the fine particles may move to the coarse region, and the coarse particle may move to the fine region, causing an uneven distribution of particles.

Several particle refinement algorithms were introduced in SPH simulations of fluid flows in recent years. A particle refinement technique for SPH was presented by Feldman and Bonet [15], by splitting one particle into several smaller particles in predefined regions. This approach is irreversible; once the coarse particles are refined into fine particles, the fine particles will not be coarsened, even if they move out the region of interest. A particle merging algorithm was introduced in SPH by Vacondio et al. [16] to coarsen particles. Another particle refinement technique was proposed by Barcarolo et al. [17], in which the coarse particles are kept after refinement such that they may be restored in the region where a high resolution is not needed. Recently, an improved particle refinement technique was proposed by Chiron et al. [18] by introducing guard particles at the coarse/fine interfaces; this method adapts the adaptive mesh refinement (AMR) approach used in the mesh-based methods to the SPH method. The particle refinement methods discussed above have a similar limitation, namely, the refinement region is a predefined and fixed region. The above methods are not applied to refinement regions that are unknown in advance, such as a violent free surface.

In the present paper, an adaptive spatial resolution (ASR) method is developed to refine the region of violent free surfaces for SPH simulations. This method is based on the adaptive refinement method proposed by Yang and Kong [19] for multiphase SPH simulations, which can adaptively refine/coarsen particles based on the distance to the interface of different phases. Nonetheless, there are several differences between the present study and the previous work [19]. First, the present study is focused on free surface flows without considering the surrounding gas, while the previous work is for multiphase flows that include both the liquid and gas phases. In multiphase flow simulations, the interface particles can be found by checking if their have neighboring particles are in a different phase. For free surface simulations, the free surface particles need to be found in a different way. Here a free surface particle searching technique is presented. Second, a particle shifting technique [20] is introduced in this work to improve particle distribution, with the consideration of variable smoothing length. Third, the variable smoothing length is improved by considering the lack of neighboring particles near the free surface.

## 2. SPH method

### 2.1 Governing equations

The Lagrangian form of the Navier-Stokes (N-S) equations are used as the governing equations of fluid.

$$\frac{d\rho}{dt} = -\rho \nabla \cdot \boldsymbol{u} \tag{1}$$

$$\frac{d\boldsymbol{u}}{dt} = \boldsymbol{g} - \frac{1}{\rho}\nabla p + \frac{\mu}{\rho}\nabla^2 \boldsymbol{u} \tag{2}$$

Here $\rho$ is density, $t$ is time, $\boldsymbol{u}$ is velocity, $p$ is pressure, $\mu$ is dynamic viscosity, and $\boldsymbol{g}$ is the gravitational acceleration.

In SPH, the governing equations can be closed by the following equation of state.

$$p = c^2(\rho - \rho_r) \tag{3}$$

Here $\rho_r$ is reference density, and $c$ is a numerical speed of sound, which is ten times larger than the maximum velocity of fluid to keep the compressibility of fluid less than 1%.

## 2.2 SPH equations

The SPH form of Eqs. (1) and (2) can be written as follows.

$$\frac{d\rho_i}{dt} = \sum_j m_j \boldsymbol{u}_{ij} \cdot \nabla_i W_{ij} \tag{4}$$

$$\frac{d\boldsymbol{u}_i}{dt} = \boldsymbol{g} - \sum_j m_j \left( \frac{p_i + p_j}{\rho_i \rho_j} + \Pi_{ij} \right) \nabla_i W_{ij} + \sum_j \frac{m_j(\mu_i + \mu_j)\boldsymbol{r}_{ij} \cdot \nabla_i W_{ij}}{\rho_i \rho_j (r_{ij}^2 + \eta)} \boldsymbol{u}_{ij} \tag{5}$$

Here $\boldsymbol{u}_{ij} = \boldsymbol{u}_i - \boldsymbol{u}_j$, $\boldsymbol{r}_{ij} = \boldsymbol{r}_i - \boldsymbol{r}_j$, $\eta = 0.01\overline{h}_{ij}^2$, and $\overline{h}_{ij} = (h_i + h_j)/2$. The subscripts $i$ and $j$ denote SPH particles, and $m$ is the mass of a particle. The summations in Eqs. (4) and (5) are taken over neighbor particles. $W = W(r, h)$ is a kernel function, $h$ is smoothing length, and $\nabla W$ is the gradient of kernel function.

The artificial viscosity proposed by Monaghan [21] is applied.

$$\Pi_{ij} = \begin{cases} \dfrac{-\alpha \overline{c}_{ij} \mu_{ij}}{\overline{\rho}_{ij}}, & \boldsymbol{u}_{ij} \cdot \boldsymbol{r}_{ij} < 0 \\ 0, & \boldsymbol{u}_{ij} \cdot \boldsymbol{r}_{ij} \geq 0 \end{cases} \tag{6}$$

Here $\alpha$ is a constant. $\overline{c}_{ij}$, $\overline{\rho}_{ij}$ and $\mu_{ij}$ are defined as follows.

$$\overline{c}_{ij} = \frac{c_i + c_j}{2}, \quad \overline{\rho}_{ij} = \frac{\rho_i + \rho_j}{2}, \quad \mu_{ij} = \frac{\overline{h}_{ij} \boldsymbol{u}_{ij} \cdot \boldsymbol{r}_{ij}}{r_{ij}^2 + \eta} \tag{7}$$

To reduce density errors, the density is reinitialized every 20 time steps using the Shephard filter [22].

$$\bar{\rho}_i = \frac{\sum_j m_j W(r_{ij}, h_i)}{\sum_j V_j W(r_{ij}, h_i)} \tag{8}$$

Here $V = m/\rho$ is the volume of a particle.

**2.3 Kernel functions**

As a meshfree method, SPH uses a kernel function to connect neighbor particles. A number of kernel functions have been used in SPH [23]. The commonly used kernels are bell-shaped functions, such as the cubic spline function.

$$W(s,h) = \alpha_d \begin{cases} (2-s)^3 - 4(1-s)^3, & 0 \leq s < 1 \\ (2-s)^3, & 1 \leq s < 2 \\ 0, & 2 \leq s \end{cases} \tag{9}$$

Here $s = r/h$, $d$ is the number of spatial dimension, and $\alpha_d$ is the normalization factor with the value of $5/(14\pi h^2)$ in two-dimensional (2D) space.

The bell-shaped kernels may result in the so-called stress instability [24, 25]. A hyperbolic kernel proposed by Yang and Liu [25, 26] can avoid the stress instability in fluid simulations, and its 2D form is as follows

$$W(s,h) = \frac{1}{3\pi h^2} \begin{cases} s^3 - 6s + 6, & 0 \leq s < 1 \\ (2-s)^3, & 1 \leq s < 2 \\ 0, & 2 \leq s \end{cases} \tag{10}$$

In the present method, the hyperbolic kernel Eq. (10) is used for Eqs. (4) and (5), while the cubic spline kernel Eq. (9) is used for Eq. (8).

**2.4 Variable smoothing length**

For SPH-USR simulations of fluid flows such as free surface flows, a constant smoothing length is usually used. However, for SPH-ASR simulations, a variable smoothing length is required because the particle distribution is not uniform and the particle spacing may change significantly.

The following variable smoothing length is used in this work,

$$h_i^{n+1} = \frac{1}{2}(\tilde{h}_i + \hat{h}_i) \quad (11)$$

where $n$ is time step, $\tilde{h}$ is a smoothing length considering the number of neighbor particles [27], and $\hat{h}$ is an average smoothing length of neighbor particles.

$$\tilde{h}_i = \frac{h_i^n}{2}\left[1 + \left(\frac{N_r}{N_i^n}\right)^{1/d}\right] \quad (12)$$

$$\hat{h}_i = \frac{1}{N_i^n}\sum_j h_j^n \quad (13)$$

Here $N_r$ is a reference constant number of particles, and $N^n$ is the number of neighbor particles at time step $n$. For free surface flows, when a particle is near the free surface, especially when a particle is leaving the free surface, Eq. (12) will result in a significant change of smoothing length. To address this numerical issue, the smoothing lengths $h^{n+1}$ and $\tilde{h}$ are only allowed to vary in the range of $0.5h_r$ to $2h_r$, where $h_r = 1.5\Delta s$ and $\Delta s = V^{1/d}$.

With variable smoothing length, the average gradient of the kernel function proposed by Hernquist and Katz [27] is applied for Eqs. (4) and (5).

$$\nabla_i \overline{W}_{ij} = \frac{1}{2}\left[\nabla_i W(r_{ij}, h_i) + \nabla_i W(r_{ij}, h_j)\right] \quad (14)$$

## 3. Particle shifting

During the simulation, SPH particles move with the fluid flow and are usually in disorder, which may reduce the accuracy of the simulation. To make the particle distribution more uniform, a particle shifting technique was introduced to SPH by Xu et al. [28], which was then generalized by Lind et al. [20] based on Fick's law. This particle shifting technique is further modified in this work for use with variable smoothing length to improve the particle distribution.

The position of a fluid SPH particle is shifted by

$$r_i^* = r_i + \delta r_i \tag{15}$$

where $\delta r$ is the particle shifting displacement vector defined as follows.

$$\delta r_i = -D_i \nabla C_i \tag{16}$$

The shifting coefficient $D$ is defined as

$$D_i = \varepsilon h_i^2 \tag{17}$$

where $\varepsilon$ is a constant with $\varepsilon \leq 0.5$ [20].

The particle concentration $C$ is defined as

$$C_i = \sum_j V_j W(r_{ij}, h_i) \tag{18}$$

The gradient of particle concentration is obtained by

$$\nabla C_i = \sum_j V_j \nabla_i W(r_{ij}, h_i) \tag{19}$$

With Eqs. (17) and (19), the particle shifting displacement vector Eq. (16) becomes

$$\delta r_i = -\varepsilon h_i^2 \sum_j V_j \nabla_i W(r_{ij}, h_i) \tag{20}$$

The particle shifting algorithm in this study differs from that proposed by Lind et al. [20] in several ways. First, the variable smoothing length is used in Eq. (20), while the original algorithm is based on the use of uniform smoothing length. Second, Lind et al. [20] modified the concentration gradient, Eq. (19), by adding an artificial pressure-like function to prevent particle pairing instability. Such a modification is not needed here because the use of the hyperbolic shaped kernel, Eq. (10), can avoid pairing instability. Third, in the present work, the particle shifting technique is only applied to particles that do not have interactions with free surface particles, because the particle shifting algorithm is based on particle diffusion and the large particle concentration gradient near the free surface can result in unexpected particle diffusion. Lind et al. [20] proposed a reduced particle shifting algorithm for the particles near the free surface. However, the reduced particle shifting algorithm may still lead to unexpected particle diffusion near a violent free surface. Thus, the particle shifting technique is not used for free surface in this work.

## 4. Search of free surface particles

In SPH simulations, the free surface is implicit rather than explicit. Thus, the free surface particles are unknown. However, the information of free surface particles is needed for the particle shifting technique presented in Section 3 and the ASR model presented in Section 5.

There are a number of algorithms to search for free surface particles [12, 29, 30]. The free surface particle searching algorithm presented in this work is similar to the arc method presented by Dilts [29] and its deformed version [31]. For a two-dimensional problem, the basic idea is to scan a circle around a candidate particle. A particle is identified as a free surface particle if there are no particles in a sector with a criticle angle.

For non-uniform particle distribution, it is not appropriate to scan a circle. In this work a reference distance is defined for scanning neighboring particles.

$$d_{ij} = C_c(\Delta s_i + \Delta s_j) \tag{21}$$

Here $C_c = 1.5$.

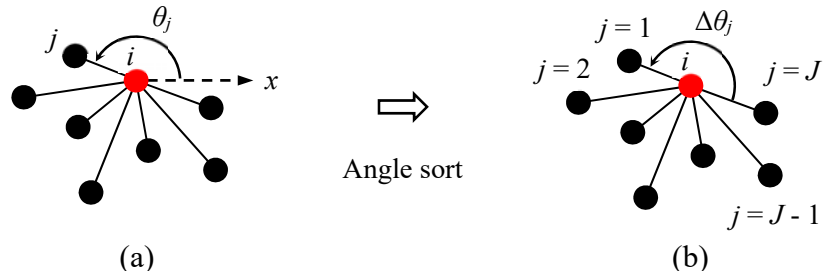

(a)　　　　　　　　　　　　　　　(b)

Fig. 1. The lines from a candidate particle to its neighboring particles and the angles between the lines.

The neighboring particles, which are usually in disorder, are sorted in a certain order before scanning. The lines from the candidate particle to its neighboring particles are used to calculate the angles from the positive $x$-direction, as shown in Fig. 1(a). Then the angles are sorted in the ascending order. Finally, the angles between adjacent lines are scanned, as shown in Fig. 1(b). If an angle is larger than the critical angle, namely, $\Delta\theta > \theta_c$, the scanning process is stopped, and the candidate particle is identified as a free surface particle. If no angle larger than the critical angle is found after the scanning process, the candidate particle is not a free surface particle. The critical angle cannot be too small or too large. If it is too large, some particles on the free surface will be identified as inner particles; if the critical angle is too small, some particles far away from the free surface may be identified as free surface particles. In the present simulations, the critical angle is set to be $\theta_c = 7\pi/18$.

Fig. 2 shows the free surface particles identified by the presented free surface searching algorithm in four cases with different surface shapes. For cases (a) and (b), the particles near the free surfaces are in order, and all the outermost particles are identified as free surface particles. For cases (c) and (d), the particles are not in order, and all the identified free surface particles are the outermost particles or the particles next to the outermost particles. It should be noted that no particles far away from the free surfaces are identified as free surface particles. This notion is very important. If any of the particles far away from free surface is identified as a free surface particle, this particle and its neighbor particles will be refined and the particle number will increase rapidly.

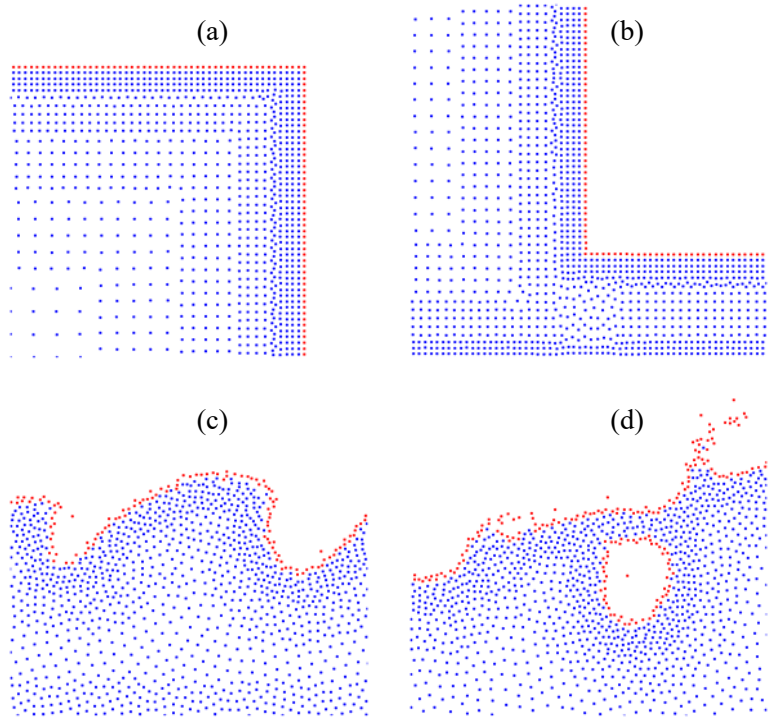

Fig. 2. Free surface particles (red) identified by the present method.

## 5. ASR algorithm

An ASR technique for free surface flows is designed to change spatial resolution adaptively with the movement of free surfaces. The present ASR technique consists two steps. The first step is to find the particles that need to be refined or coarsened. The second step is to refine the particles by particle splitting or coarsen the particles by particle merging.

For ASR, both particle spacing and particle mass are not uniform. In the present ASR method, particle spacing is used to determine the spatial resolution, and particle mass is used to check whether a particle needs refinement or coarsening. If the mass of a particle is larger than the critical mass for splitting, the particle will be refined. On the contrary, if the mass of a particle is less than the critical mass for merging, the corresponding particles will be coarsened.

### 5.1 Reference particle spacing

The reference particle spacing is associated with the particle band. As shown in Fig. 3, the particles are divided into a series of particle bands that are parallel to the free surface. The particles in the same band have the same reference particle spacing. With the increase of distance to the free surface, both the reference particle spacing and the width of the band are increased. The width of the particle band is set to $K\Delta S$, where $\Delta S$ is the reference particle spacing and $K$ is the ratio of the kernel width to the particle spacing. The reference particle spacing increases with the distance to free surface,

$$\Delta S_{k+1} = C_r \Delta S_k \tag{22}$$

where $k$ (= 0, 1, 2, …) is the index of particle band and increases with the increase in distance to the free surface. The parameter $C_r$ is defined as the adaptive number, which is the ratio of the reference particle spacing of two adjacent particle bands.

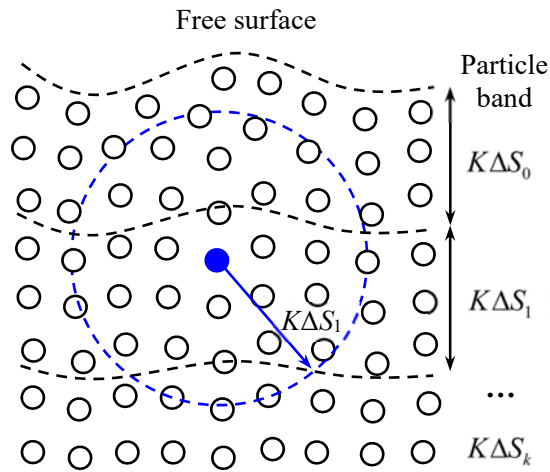

Fig. 3. Schematic of particle band and reference particle spacing.

**5.2 Particle splitting and merging**

To determine whether the particle mass is large enough for splitting or small enough for merging, a reference mass $m_r$ is defined for each particle.

$$m_r = \rho_r \Delta S^d \tag{23}$$

The ratio of the mass of a particle to its reference mass is $\gamma = m/m_r$. When the mass ratio $\gamma$ is larger than the critical value for splitting $\gamma_s$, namely, $\gamma > \gamma_s$, the particle is split into two smaller particles, as shown in Fig. 4(a). When the mass ratio $\gamma$ is less than the critical value for merging $\gamma_m$, namely, $\gamma < \gamma_m$, the particle is merged with its nearest neighboring particle, as shown in Fig. 4(b). Note that for both particle splitting and merging in Fig. 4 the particle $k$ is the nearest particle of particle $i$. According to Yang and Kong [19], $\gamma_s = 1.5$ and $\gamma_m = 0.7$.

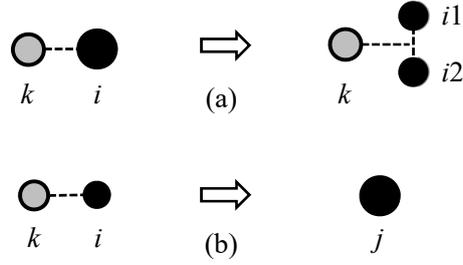

Fig. 4. Schematic of (a) particle splitting and (b) particle merging.

With the consideration of mass and momentum conservations, the mass and velocity of the two small particles after splitting can be obtained as follows.

$$m_{i1} = m_{i2} = \frac{m_i}{2} \quad (24)$$

$$\boldsymbol{u}_{i1} = \boldsymbol{u}_{i2} = \boldsymbol{u}_i \quad (25)$$

The positions of the two particles are calculated by

$$\boldsymbol{r}_{i1} = \boldsymbol{r}_i + \frac{\lambda}{2} \Delta s_i \boldsymbol{e}_{ik}^{\perp} \quad (26)$$

$$\boldsymbol{r}_{i2} = \boldsymbol{r}_i - \frac{\lambda}{2} \Delta s_i \boldsymbol{e}_{ik}^{\perp} \quad (27)$$

where $\boldsymbol{e}_{ik}^{\perp}$ denotes the unit vector perpendicular to the line connecting particles $i$ and $k$, and $\lambda = 0.6$ is the ratio of the particle spacing after splitting to the particle spacing before splitting [19].

The properties of the particle after particle merging are obtained as follows.

$$m_j = m_i + m_k \quad (28)$$

$$\boldsymbol{u}_j = \frac{m_i \boldsymbol{u}_i + m_k \boldsymbol{u}_k}{m_i + m_k} \quad (29)$$

$$r_j = \frac{m_i r_i + m_k r_k}{m_i + m_k} \tag{30}$$

## 6. Numerical examples

In this section, the SPH-ASR method is validated by simulating various cases of violent free surface flows. The SPH-ASR results are compared to SPH-USR results and experimental data from literatures.

In the following examples, the ASR algorithm is applied to fluid particles only. The resolution of the wall particles remain the same to their initial resolution. The resolution of the fluid particles near the solid walls varies adaptively with their distances to the solid walls using the ASR algorithm.

### 6.1 Dam-break flow on a dry bed

The schematic of the dam-break flow on a dry bed is shown in Fig. 5, with a configuration similar to that in an experiment [32]. The water column on the left side of the tank is initially static. The water will flow to the right side of the tank due to gravitation and impact the right wall of the tank. Fluid pressure is measured at point $P_1$ on the right wall.

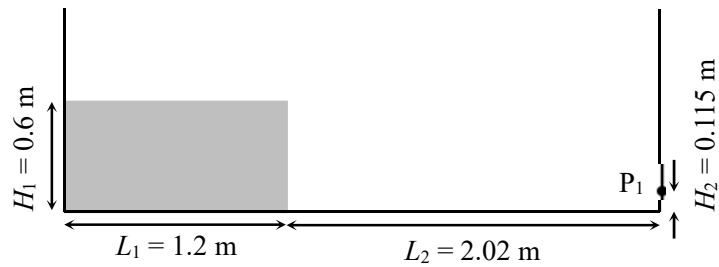

Fig.5. Configuration of the dam-break flow on dry bed. Fluid pressure is measured at point $P_1$.

Table 1 shows the particle spacing, particle number and computational time of SPH simulations using different spatial resolutions. Both the particle number and CPU time of the ASR case are lower than those of the USR case. The particle spacing of ASR changes with space and time, resulting in the change of the number of particles.

To compare the particle number of different spatial resolutions, a relative particle number is defined as the ratio of the ASR particle number to the USR particle number. Fig. 6 shows the relative particle number as a function of time. The change of the particle number of ASR is due to the flow of water and movement of the free surface, as shown in Fig. 7. It can be seen from Fig. 7 that the particles near the free surface and solid wall have the finest spatial resolution, while the particle spacing increases with the distance from the free surface and solid wall. When the water flows to the right, the large particles on the left come closer to the free surface and they are refined adaptively. When enough particles gather on the right, the particles far away from the free surface are coarsened adaptively.

Table 1. Particle spacing, particle number and CPU time of SPH-USR and SPH-ASR simulations of dam break flow on a dry bed.

|  | Particle spacing (mm) | Particle number | CPU time (min) |
| --- | --- | --- | --- |
| SPH-USR | 10 | 9,624 | 47 |
| SPH-ASR | 10 ~ 28 | 5,876 ~ 8,957 | 43 |

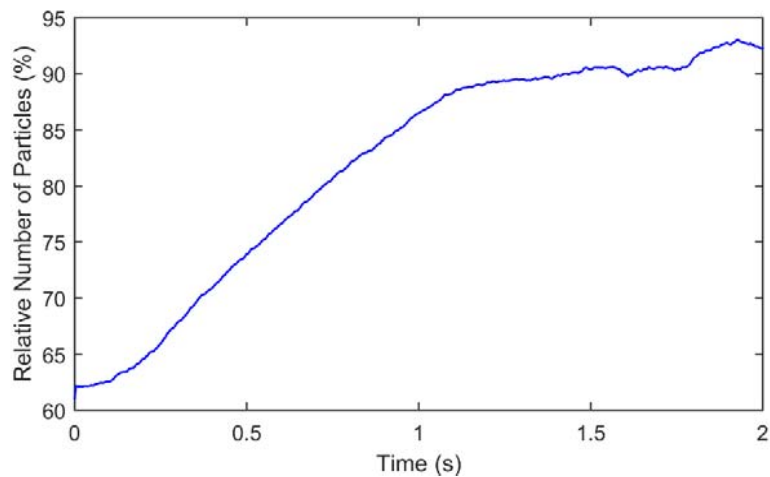

Fig. 6. Relative number of particles as a function of time.

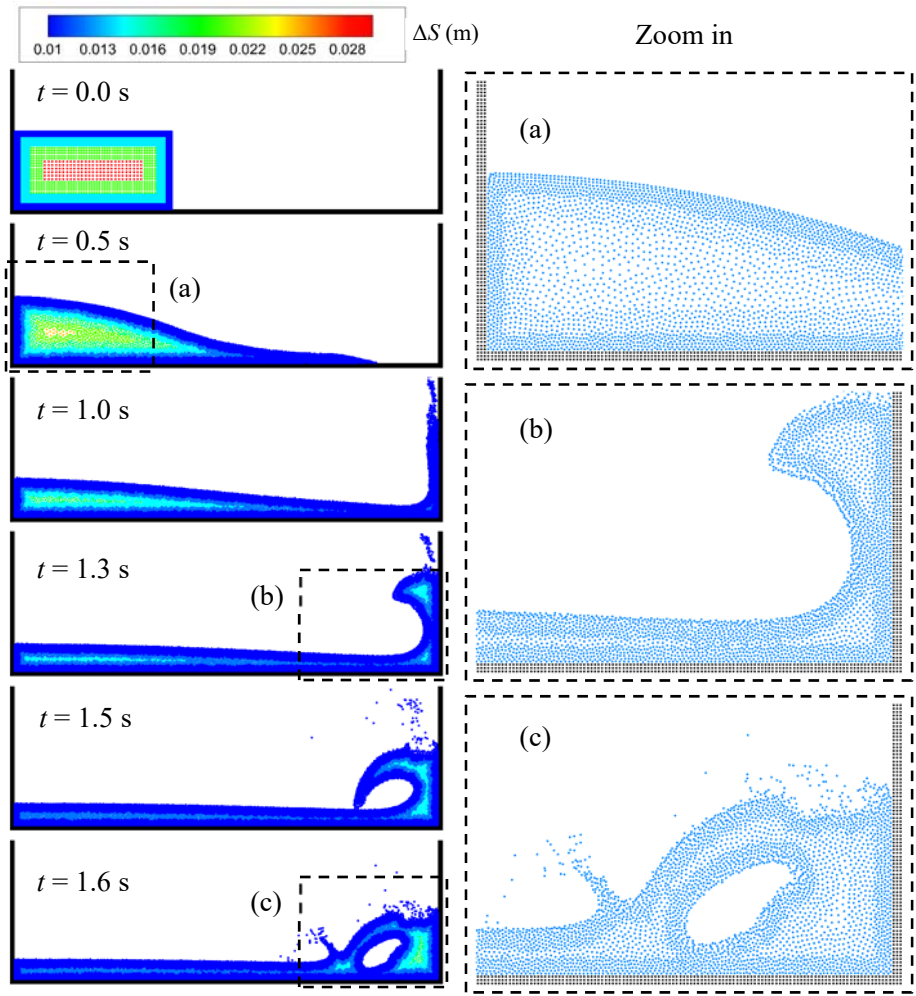

Fig. 7. Evolution of particle spacing of the dam-break flow on a dry bed from SPH-ASR simulation.

Fig. 8 compares the evolutions of the free surface and pressure distribution obtained using SPH-ASR and SPH-USR. The numerical simulations using the two methods are in good agreement, although the spatial resolutions are different. A quantitative comparison of the pressure at point $P_1$ between experiment [32], SPH-USR and SPH-ASR is shown in Fig. 9. The fluid reaches point $P_1$ around $t = 0.6$ s. Note that there was a noticeable pressure fluctuation before $t = 0.6$ s in the experimental data; this fluctuation does not appear to be realistic because the fluid has not arrived yet. Despite such an experimental uncertainty, the present SPH method performs well in predicting this dam-break flow.

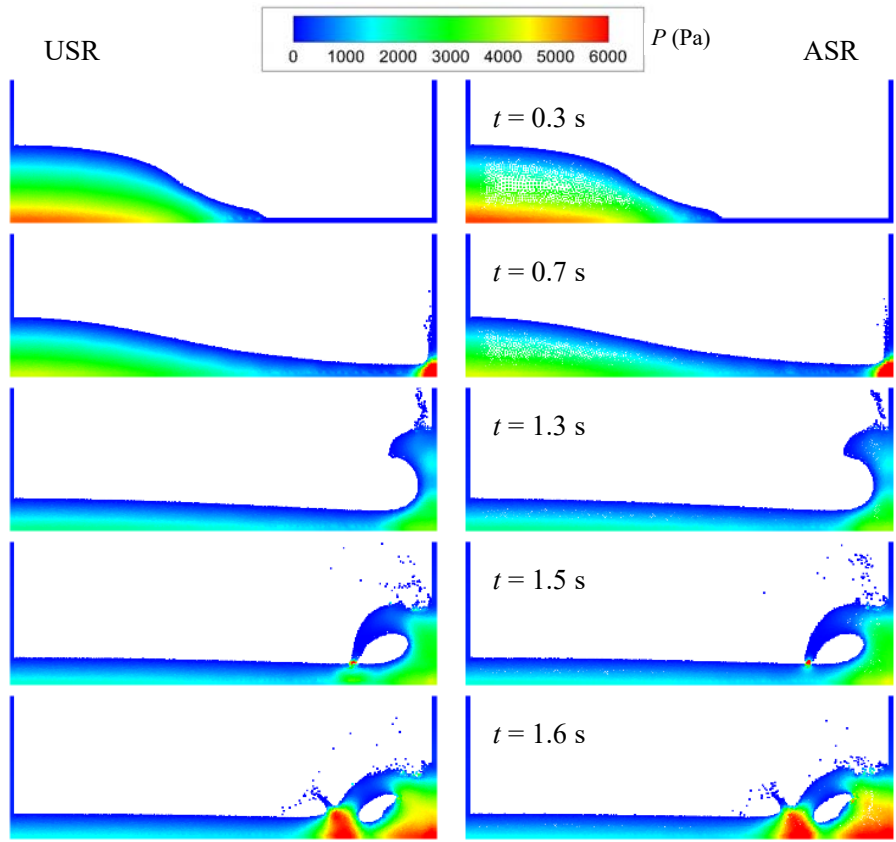

Fig. 8. Comparison of the free surface and pressure distribution between SPH-USR (left) and SPH-ASR (right).

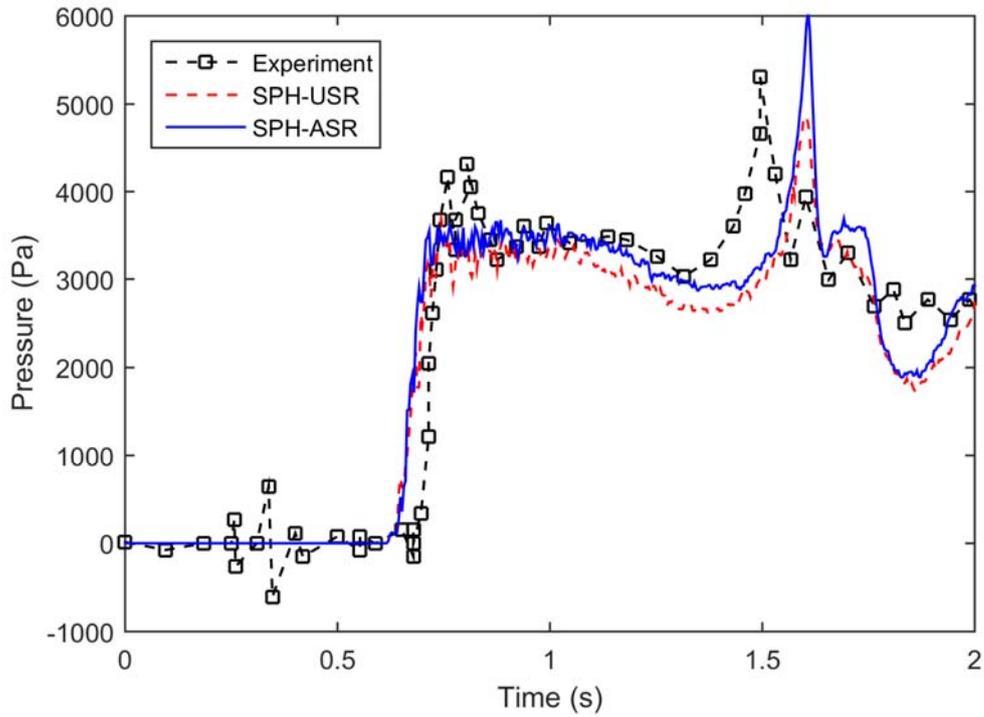

Fig. 9. Comparison of the pressure at $P_1$ between experiment [32], SPH-USR and SPH-ASR.

**6.2 Dam-break flow on a wet bed**

The dam-break flow on a wet bed is more complex than the dam-break flow on a dry bed, because of the consideration of the effects of gate movement. Fig. 10 shows the initial computational setting of the dam-break flow on a wet bed, as studied in the experiment [33]. Here the velocity of the gate is set to 1 m/s, while the gate velocity of the experiment is unknown. When the gate moves up, the deep water flows to the wet bed on the right. Table 2 shows the particle spacing, particle number and CPU time of SPH-ASR and SPH-USR simulations. It can be seen that the ASR simulation can reduce the particle number and CPU time significantly compared to the USR simulation.

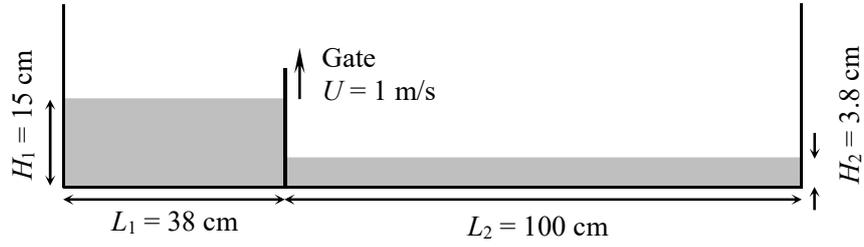

Fig. 10. Configuration of dam-break flow on a wet bed.

Table 2. Particle spacing, particle number and CPU time of SPH-USR and SPH-ASR simulations of dam break flow on a wet bed.

|  | Particle spacing (mm) | Particle number | CPU time (min) |
| --- | --- | --- | --- |
| SPH-USR | 2 | 26,906 | 201 |
| SPH-ASR | 2 ~ 8 | 16,056 ~ 17,788 | 126 |

Fig. 11 shows the evolution of the particle spacing of the dam-break flow on a wet bed from SPH-ASR simulation. It can be seen that the particle spacing changes adaptively with the movement of the free surface.

Fig. 12 compares the SPH-ASR and SPH-USR results to the experimental data [33]. The SPH-ASR and SPH-USR simulations are in good agreement on both the pressure distribution and the shape of the free surface. However, the free surface predicted by SPH moves about 0.065 second faster than the experiment, possibly because of different starting time.

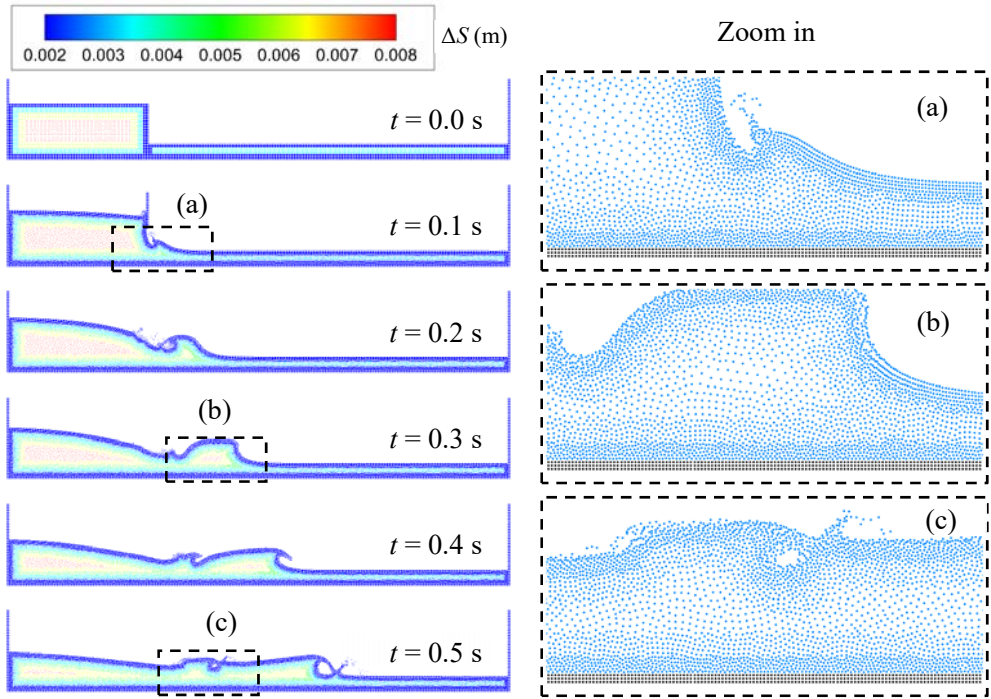

Fig. 11. Evolution of particle spacing of the dam-break flow on a wet bed by SPH-ASR.

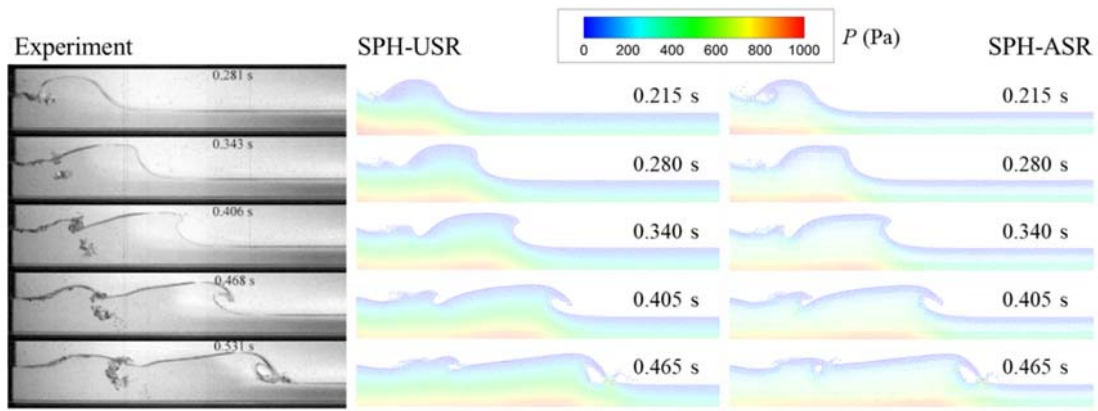

Fig. 12. Comparison of the dam-break flow on a wet bed between experiment (left) [33], SPH-USR (middle) and SPH-ASR (right).

**6.3 Water entry of a horizontal cylinder**

Fig. 13 shows the schematic of water entry of a horizontal cylinder. The density of the cylinder is half the water density. The information of particle spacing, particle number and CPU time is shown in Table 3. Note that in the previous two cases, the particle numbers used for ASR and USR are in the same order. In this case, the particle number used for ASR is less than 10% of the particle number used for USR. The computational time of the ASR simulation is 10.5% of the time of the USR simulation.

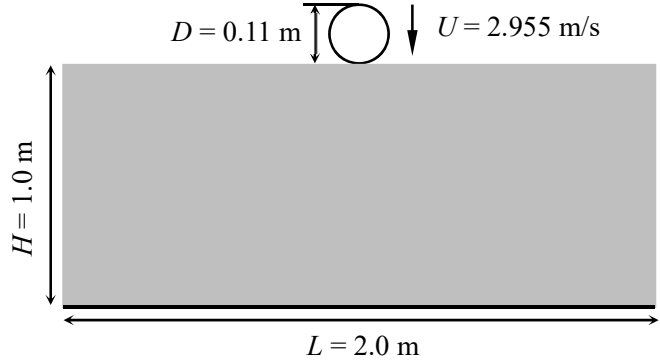

Fig. 13. Configuration of water entry of a horizontal cylinder.

Table 3. Particle spacing, particle number and CPU time for different spatial resolutions of SPH simulations of water entry of a horizontal cylinder.

| Resolution | Particle spacing (mm) | Particle number | CPU time (min) |
| --- | --- | --- | --- |
| SPH-USR | 5 | 81,467 | 38 |
| SPH-ASR | 5 ~ 40 | 7,092 ~ 8,145 | 4 |

The SPH-ASR simulation gives very similar results to the SPH-USR simulation with a significantly reduced particle number and CPU time, as shown in Fig. 14. The penetration depth of the cylinder in water obtained using SPH-ASR also agrees well with the results using SPH-USR and the experimental data [34], as shown in Fig. 15.

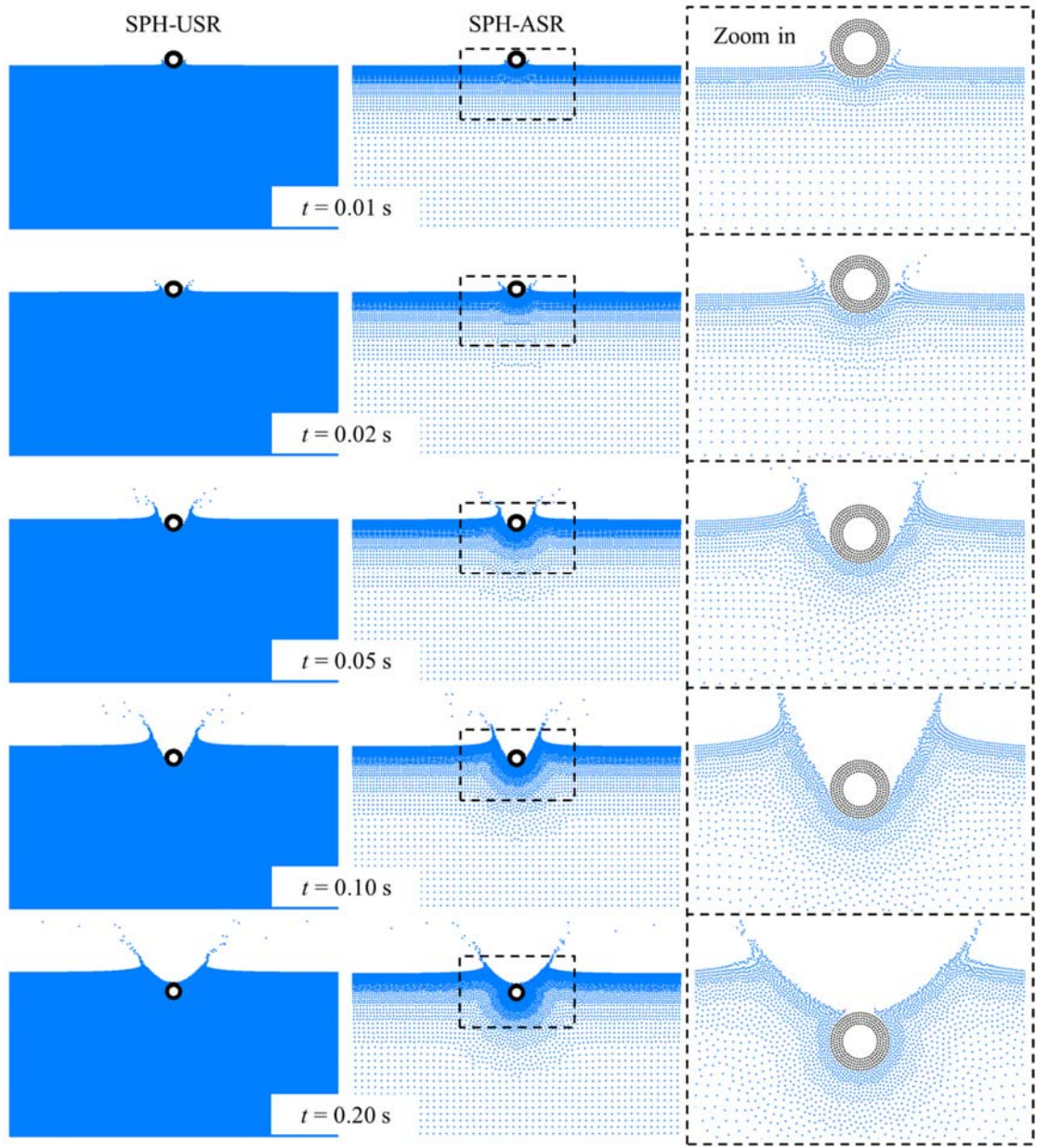

Fig. 14. Comparison of water entry of a horizontal cylinder using SPH-USR (left), SPH-ASR (middle) and zoom in of the cylinder region (right).

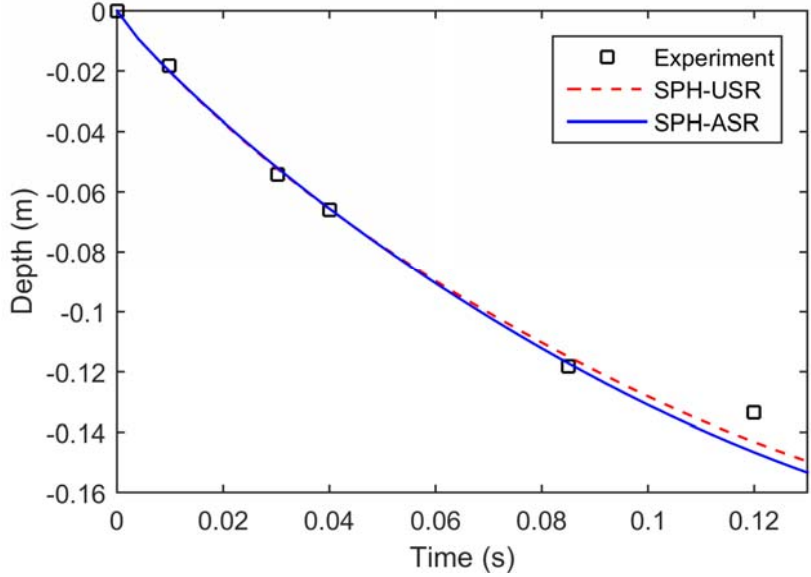

Fig. 15. Comparison of penetration depth of cylinder in water between experiment [34], SPH-USR and SPH-ASR simulations.

**6.4 Water entry of a slender body**

The configuration of water entry of a slender body is similar to that in Fig. 13. The width and height of the slender body are 0.01 m and 0.1 m, respectively. The initial velocity is 5 m/s. The density of the slender body is 9 times the water density. In comparison to the cylinder case in Section 6.3, this case uses a high spatial resolution. As shown in Table 4, the high spatial resolution results in a large number of particles and a high computational cost. For the SPH-USR simulation, the particle number is about 2 million, and the CPU time is about 81 hours. In the SPH-ASR simulation, the average particle number is less than 0.05 million, and the CPU time is only 1.85 hours.

Table 4. Particle spacing, particle number and CPU time for different SPH simulations of water entry of a slender body.

| Resolution | Particle spacing (mm) | Particle number | CPU time (h) |
| --- | --- | --- | --- |
| SPH-USR | 1 | 2,006,624 | 81.2 |
| SPH-ASR | 1 ~ 32 | 36,623 ~ 50,025 | 1.85 |

Fig. 16 compares the simulation results using SPH-USR and SPH-ASR. After the slender body impacts the water surface, the water near the front of the body is pushed away and forms a splash and a cavity. Then the cavity is closed by the water on the top. Finally, the cavity shrinks because of the surrounding water and even breaks up into a few tiny cavities. This process is captured by both the SPH-USR and SPH-ASR simulations, while the computational cost of SPH-ASR is much less than SPH-USR.

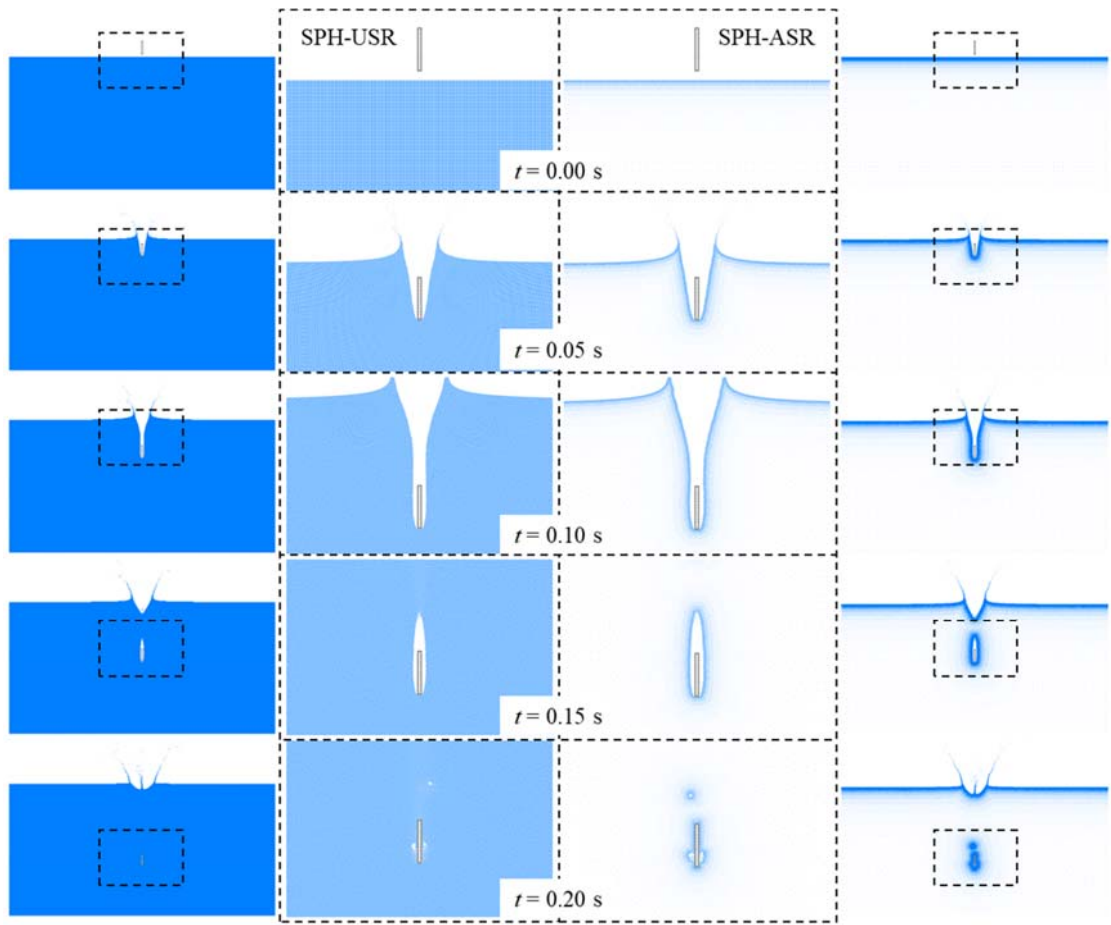

Fig. 16. Comparison of water entry of a slender body between SPH-USR (left) and SPH-ASR (right) simulations. The two middle columns are zoom-in of regions around the body.

## 7. Conclusion

An SPH-ASR method for simulating free surface flows was presented in this paper. The spatial resolution changes adaptively with the movement of the free surface. Particle splitting and merging techniques were developed for particle refinement and coarsening, respectively. The present method uses a variable smoothing length because of the variation in particle spacing during the SPH simulations. A particle shifting technique was introduced to improve particle distribution.

Various free surface flow problems were studied to validate the present SPH-ASR method. The numerical results show that the ASR method is able to adaptively change the particle spacing according to the distance to the free surface, even if the free surface moves violently. Good levels of agreement were obtained between the SPH-ASR results, SPH-USR results and experimental data.

The efficiency of ASR depends on the levels of spatial resolution, namely, the finest particle spacing and the proportion of the refinement area. For the two dam-break flow problems, the particle spacing varies in a small range and most of the area is refined, thus the use of ASR moderately reduces the number of particles and computational time. For the two water entry problems, the particle spacing varies in a large range and the proportion of the largest particles is high, thus the use of ASR significantly reduces the number of particles and computational time.

**Acknowledgement**

This work was supported by the National Natural Science Foundation of China (Grant No. 11872117).